# Singular Values using Cholesky Decomposition

Aravindh Krishnamoorthy, Kenan Kocagoez
ST-Ericsson AT GmbH, Nuernberg, http://www.stericsson.com
{aravindh.k, kenan.kocagoez}@stericsson.com

*Abstract*—In this paper two ways to compute singular values are presented which use Cholesky decomposition as their basic operation.

## I. Introduction

Cholesky Decomposition is a fast and numerically stable matrix operation that finds wide use in various fields for linear system solving, inversion, and factorisation [1] and is a common block for matrix operations in hardware and software based computing systems. In this paper we present two ways of finding singular values of a matrix which use Cholesky decomposition as their basic operation, and may be useful in constrained computing environments such as embedded computing systems.

The first method "Cholesky Iterations" uses Cholesky decompositions based iterations [2], and [10]. The second method uses Cholesky decomposition to compute the QR factors, which are further used for "QR Iterations" [1], [4], [5], and [6].

### A. Cholesky Decomposition

Cholesky Decomposition [1] factorises a complex (or real-valued) positive-definite Hermitian symmetric matrix into a product of a lower triangular matrix and its hermitian transpose.

$$A = R^H R, \text{ or equivalently, } R = \mathrm{chol}(A)$$

where $A \in C^{n \times n}$ is a positive-definite Hermitian symmetric matrix, $R \in C^{n \times n}$ is an upper triangular matrix, and $R^H$ is the Hermitian tranpose of $R$.

### B. For Positive Semi-definite Matrices

Cholesky decomposition for positive semi-definite matrices is analysed in [3], and it is computed easily by setting the row to zero if the diagonal element is found to be zero.

**Algorithm 1** Cholesky Decomposition for Positive Semi-definite Matrix

$\quad$ **for** $i = 1...N$ **do**
$\quad\quad R_{ii} = \sqrt{A_{ii} - \sum_{k=1}^{i-1} R_{ki} R_{ki}^*}$
$\quad\quad$ **if** $|R_{ii}| > \epsilon$ **then**
$\quad\quad\quad$ **for** $j = i+1...N$ **do**
$\quad\quad\quad\quad R_{ij} = \frac{1}{R_{ii}} \left( A_{ij} - \sum_{k=1}^{i-1} R_{kj} R_{ki}^* \right)$
$\quad\quad\quad$ **end for**
$\quad\quad$ **else**
$\quad\quad\quad R_{i,i+1...N} = \mathbf{0}$
$\quad\quad$ **end if**
$\quad$ **end for**

In the algorithm above, instead of comparing the diagonal element with zero, we compare it against a small tolerance value $\epsilon$ to account for numerical inaccuracies.

23rd January, 2012

## II. Cholesky Iterations

In this section, "Cholesky Iterations" based computation of singular-values is discussed for positive (semi-) definite, symmetric and arbitrary matrices. While the performance for positive (semi-) definite and symmetric matrices are acceptable, the method for arbitrary matrices may have poor performance for ill-conditioned matrices.

### A. Positive (semi-) definite Matrices

Let $A \in C^{n \times n}$ be a positive (semi-) definite Hermitian symmetric matrix, then the singular values (also eigenvalues) are given using the following algorithm.

**Algorithm 2** Cholesky Iterations - Positive (semi-) definite matrices

$\quad J^{(0)} = A$
$\quad$ **for** $k := 1...\mathrm{iter}$ **do**
$\quad\quad R^{(k)} = \mathrm{chol}(J^{(k-1)})$
$\quad\quad J^{(k)} = R^{(k)} (R^{(k)})^H$
$\quad$ **end for**
$\quad \Sigma = \Lambda = \mathrm{diag}(J^{(\mathrm{iter})})$

where

$$\mathrm{diag}(A) = \begin{cases} A_{ii} & \text{if i = j} \\ 0 & \text{if i} \neq \text{j} \end{cases}$$

Singular values lie along the diagonals of the matrix $\Sigma$, and they are the same as eigenvalues.

*1) Diagonalisation:*
If
$$J^{(k-1)} = R^H R$$
then
$$R J^{(k-1)} R^{-1} = R R^H = J^{(k)}$$
and so the matrix is diagonalised as:
$$J^{(k)} = (R^{(k-1)}...R^{(1)}) J^{(0)} (R^{(k-1)}...R^{(1)})^{-1}$$

The matrix product $(R^{(k-1)}...R^{(1)})$ increases unboundedly with iterations, but the values of $J^{(k)}$ converge towards the diagonal singular-values matrix.

## B. Symmetric matrices

Let $A \in C^{n \times n}$ be a Hermitian symmetric matrix, then the singular values (and eigenvalues) are given using the following algorithm.

**Algorithm 3** Cholesky Iterations - Symmetric Matrices
---
$J^{(0)} = A + \mu I$
**for** $k := 1...\text{iter}$ **do**
$\quad R^{(k)} = \text{chol}(J^{(k-1)})$
$\quad J^{(k)} = R^{(k)}(R^{(k)})^H$
**end for**
$\Lambda = \text{diag}(J^{(\text{iter})}) - \mu I$
$\Sigma = |\Lambda|$

---

Where $\mu$ is a pivoting constant to ensure that the matrix $J^{(0)}$ is positive (semi-) definite matrix; $\mu$ may be found using variety of pivoting techniques, for e.g. see [7], [8], and [9].

If the symmetric matrix is well-conditioned, the algorithm given below for Aribtrary Matrices may also be used, which avoids the computation of the pivoting constants.

## C. Arbitrary Matrices

Let $A \in C^{m \times n}$ be an arbitrary matrix, then the singular values are given using the following algorithm.

**Algorithm 4** Cholesky Iterations - Arbitrary Matrices
---
$J^{(0)} = A^H A$
**for** $k := 1...\text{iter}$ **do**
$\quad R^{(k)} = \text{chol}(J^{(k-1)})$
$\quad J^{(k)} = R^{(k)}(R^{(k)})^H$
**end for**
$\Sigma = \sqrt{\text{diag}(J^{(\text{iter})})}$ (or) $\text{diag}(R^{(\text{iter})})$

---

If the matrix $A$ is full row-rank, $J^{(0)} = AA^H$ may also be used as an initial state for the iterations.

## III. QR Iterations

In this section, Cholesky decomposition is used to compute the QR factors of an aribtrary matrix. These QR factors are further used for "QR Iterations" to compute the singular values.

While computationally the method of finding QR factors from Cholesky decomposition may be expensive, its performance is acceptable and it may be advantageous to use this method in certain constrained computational platforms.

If $A \in C^{m \times n}$ is an arbitrary matrix, then the QR Decomposition is given by:

$$R = \text{chol}(A^H A) \quad (1)$$
$$Q = AR^{-1} \quad (2)$$

If $A$ is not full-column rank, we may find $Q$ as $Q = AR^+$ where $R^+$ is the pseudo-inverse of $R$. In that case, the last $m - n$ columns of $Q$, and the lower $m - n$ rows of $R$ are zeros and may be ignored.

**Theorem III.1.** *Matrix $Q = AR^{-1}$ is Unitary.*

Proof: We have $A^H A = R^H R$; pre-and-post multiplying with $R^{-H}$ and $R^{-1}$, we have, $R^{-H} A^H A R^{-1} = I$ and so $(AR^{-1})^H AR^{-1} = I$, or $Q^H Q = I$, and hence $Q$ is unitary.

**Algorithm 5** (Pure) QR Iterations
---
$J^{(0)} = A$
**for** $k := 1...\text{iter}$ **do**
$\quad R^{(k)} = \text{chol}((J^{(k-1)})^H J^{(k-1)})$
$\quad Q^{(k)} = J^{(k-1)}(R^{(k)})^{-1}$
$\quad J^{(k)} = R^{(k)} Q^{(k)}$
**end for**
$\Sigma = \text{diag}(J^{(\text{iter})})$

---

For further information on "QR Iterations" refer [1], [4], [5], and [6].